# Multi-Level Mesa

By Thomas Pike
PhD Candidate, Computational Social Science
George Mason University
tpike3@gmu.edu

**Abstract:** Multi-level Mesa is an extension to support the Python based Agents Based Model (ABM) library Mesa. Multi-level Mesa provides ABM infrastructure to allow for the inclusion of complex networks, which have modules (groups) and hierarchies (layers) of agents. This approach allows for users to define and simulate multi-layered adaptions of complex networks. This study reviews other multi-level libraries currently in the field, describes the main functions and classes of the Multi-level Mesa, and describes its implementation and impact in numerous varieties using the seminal ABM - Sugarscape. Multi-level Mesa and Sugarscape examples are available on GitHub at https://github.com/tpike3/multilevel_mesa and https://github.com/tpike3/SugarScape.

Multi-Level Mesa provides ABM infrastructure to support modules and hierarchies. Modularity is the concept that clusters of linked nodes within a network can effectively act as a single node, and can also be known as communities or building blocks (Barabasi, 2016; Holland, 1995). Hierarchies represent the layers of emergence which can occur within a complex system. For example, using an individual as the focal point, hierarchies go smaller from the individual, as a human body is comprised of organs which are comprised of cells, which are comprised of chemicals and so on. Or, larger from the individual, as humans make up families, which make up neighborhoods, which make up towns and so on (Miller & Page, 2007). Modules and hierarchies are a critical part of complex adaptive systems as they provide sub-assemblies to retain working systems while enabling adaption (Holland, 1995; Simon, 1997). As an essential aspect of complex systems modules and hierarchies are critical to analyze.

Object Oriented Programming (OOP) inherently consists of modularity and hierarchies and exploiting this capability is what provides greatest advantage to OOP languages (Booch et al., 2007). OOP provides analysts the ability to capture specific modules, hierarchies and processes of complex adaptive systems. ABM platforms and coding libraries then exploit the properties of OOP by providing ABM infrastructure. This infrastructure reduces the cost of the modeler who can focus on simulating his or her phenomenon of interest and not on writing code which manages the interactions of the phenomenon. ABMs, however, typically stop at two levels of interaction. Agents produce the bottom-up emergent behavior of the next level, but no further hierarchies are produced (Haman Tchappi, Galland, Kamla, & Kamgang, 2018; Morvan, 2013). Multi-level Mesa seeks to further extend the typical ABM dynamic to enable more complex interactions where agents and groups of agents can interact across multiple hierarchies and have cascading effects across those hierarchies. The goal of Multi-level Mesa is to provide methods to help manage the complex interactions of agents and modules of agents (e.g. groups) across multiple hierarchies (e.g. levels).

Multi-level Mesa starts from the view of complex systems as adaptive networks and allows not only for the formation and dissolution of modules but also for active and resting modules (or neutral networks) which can interact across layers. In the taxonomy of efforts to facilitate multiple levels in ABMs, Multi-level Mesa falls in the category of generalizable coding libraries (Taillandier et al. 2012; Morvan, 2013) and is the only Python based library. Multi-level Mesa is intended to be a readily available coding library which can be employed to support models developed to understand adaptive networks.

This chapter proceeds in four sections. First, a literature review of current multi-level ABM approaches. Second, a discussion of Multi-level Mesa's conceptual approach. Third, a discussion of Multi-level Mesa's methods. Fourth, an implementation of Multi-level Mesa using the Sugarscape construct developed by Rob Axtell and Joshua Epstein (1996).

## Literature Review

As modules and hierarchies are an inherent feature of complex systems there is a rich body of literature examining them across multiple disciplines. This expansive body of literature can be broken down into three broad categories. First, are approaches which want to identify existing processes that produce emergent modules, which in turn reify and become agents at a higher hierarchy. Second, are approaches which provide computational infrastructure so others can dictate their own emergent and reification processes. Third, are attempts at linking different models together each of which is its own module and falls within certain hierarchies. Due to the wide breadth of research a complete review is impractical, instead this survey will provide a brief overview of the first category, as it has the largest amount of literature, and then focus on the existing computational infrastructure before discussing attempts to link models together. To begin, however, it is important to discuss some of the various terms. Due to the large amount of research across multiple disciplines on this subject a diverse terminology has emerged for the concept of modules and hierarchies within a complex system.

**Terminology**

The three main terms for deliberate inclusion of modules and hierarchies into ABMs are multi-level, multi-scale and holons. The term multi-scale is favored by the natural sciences but contentious to other disciplines who argue multi-scale is inaccurate (Gil-Quijano, Louail, and Hutzler 2012; Morvan, 2013). Two cities, for example, Tucson, Arizona and New York, New York are both at the same 'city-level', but are of two different population scales, 535,000 and 8.6 million, respectively ("Population in the U.S. - Google Public Data Explorer," 2018). Natural sciences may counter families and states are really just different scales of human organization in one level of the Earth's ecological hierarchy, so the proper terminology is really determined by one's perspective. The second term is holon, which has accompanying descriptors such as holarchy, for discussing the hierarchies within the system, or holonic, to describe a system with modules and hierarchies (Haman Tchappi et al., 2018). Holon comes from Arthur Koestler's book *Ghost in the Machine* and was invented to specifically address the existence of subassemblies within complex systems based on Herbert Simon's parable of the two watchmakers (1967; 1997). The coding module adopts the term multi-level, as it is more descriptive for the user who is concerned about the levels within a specific field. Multi-level was then selected instead of holon for the simple reason the term multi-level makes its purpose more obvious to potential users. Despite these different terms, multi-scale, multi-level and holon each mean the deliberate inclusion of modules and hierarchies.

There are two other terms worthy of discussion in the literature, which intersect with multi-level ABMs but also have models outside the ABM set. First is multi-modelling (also referred to as meta models). This effort can be seen as a separate but intersecting focus area. Multi-models are an effort to link two or more models of a similar phenomenon together to allow for numerous research efforts to be combined (Scerri, Drogoul, Hickmott, & Padgham, 2010; Soyez, Morvan, Dupont, & Merzouki, 2013). Some of these efforts fall under the third category discussed in this literature review, while some are wholly independent from ABMs, notably the Coupled Earth System Model which consists of four publicly available models to explore the Earth's weather system.[1] The second term is hybrid ABMs, these are primarily system biology models and combine ABMs with systems dynamics where one or more levels is agent based and their actions parameterize differential equations at other levels, whose output provides inputs to the agents (Cilfone, Kirschner, & Linderman, 2015; Smallwood & Holcombe, 2006). Each of these research areas also examine complex systems at multiple hierarchies, but are specialized approaches which can be seen as overlapping. With the main terminology described, the next step is to review each of the three categories of research efforts.

**Category One: Processes**

The first category is research trying to discover or define processes for creating multiple hierarchies. This category consists of the largest amount of research and exists across multiple fields. This research tries to address the theoretical issue of generalizable mechanisms for identifying and reifying emergent phenomenon and cross-level communication (Haman Tchappi et al., 2018; Morvan, 2013; Seck & Honig, 2012). For example, when a group of bacteria form a microbial colony and begin to act as a singular entity, a group of cells form a functioning organ, or a population of people act as a single nation. The natural disciplines prefer the term multi-scale and have well developed and coordinated research efforts to try to identify emergence and reification processes, which includes government sponsored projects, working groups, tools, databases, webinars and competitions, coding platforms and modeling languages (Falcone, Chopard, and Hoekstra 2010; "Interagency Modeling and Analysis Group" 2018; Morvan, 2013; Smallwood and

---

[1] The climate models are available at http://www.cesm.ucar.edu/models/ccsm4.0/

Holcombe 2006).[2] Ecology has a series of models which look at the dynamics of multiple levels within a trophic web using ABMs. Existing ecological models focus on scale (level) transfer or clustering methods to explore the dynamics of how agents coalesce and how their actions impact levels above and below them from the micro level (e.g. soil) to the macro level (e.g. an ecosystem) (Morvan, 2013). The natural sciences have extensive work examining multiple levels of complex systems with significant effort placed on understanding the emergence and reification of entities, and interdependencies between levels.

Expanding beyond biology and ecology, multi-level models are represented in three areas. The first area is traffic and pedestrian models. For these models, levels are added to make the models more computationally efficient as pedestrians begin to move together as a type of flocking model (Haman Tchappi et al., 2018; Navarro, Corruble, Flacher, & Zucker, 2013). The second area overlaps with traffic and pedestrian models to examine city development. This area includes several research efforts which try and address different aspects of the multi-level problem. These aspects include identifying when new agents emerge (Camus, Bourjot, & Chevrier, 2013; Gil-Quijano et al., 2012) and how different levels and modules should interact with each other, which intersects with research into coupling models together (e.g. multi-models) (Navarro et al., 2013). The third area is organizational and has seen applications trying to manage intelligent autonomous intelligence vehicles. In this area multiple levels are used to find ways to deconflict layers within an organization, such as fleets of vehicles autonomously conducting port operations (Haman Tchappi et al., 2018; Soyez et al., 2013). These three areas show considerable cross-fertilization as they are looking at similar systems of flow and organization for different purposes.

Due to the cross fertilization of the previous three areas there are general frameworks which are used and improved upon for their specific research problems. These frameworks are CRIO (Capacity, Role, Interaction, Organization) (Haman Tchappi et al., 2018) , IRM4MLS (Influence Reaction Model for Multi-Level Simulations) (Soyez et al., 2013), and AA4MM (Agents and Artifacts for Multi-Modelling) (Camus et al., 2013; Siebert, Ciarletta, & Chevrier, 2010). Interestingly, each of these approaches are proposed by French universities, who have the most research papers, outside the natural sciences, on this subject.

Research into natural processes for the emergence and reification of new layers and the cross communication between layers represented the largest amount of research on multi-level ABMs. The natural sciences have the most developed research efforts to examine this problem. There is also substantial research examining these phenomena in population flow and organizational models. Although this review focused on ABMs, there are similar efforts in discrete event simulations, specifically, the DEVs models, whose evolution over time has made them more similar to ABMs (Haman Tchappi et al., 2018; Morvan, 2013; Seck & Honig, 2012)

---

[2] A concise website containing multi-scale modelling efforts and links to models, tools and databases is located at https://www.imagwiki.nibib.nih.gov/.

**Category Two: Computational Infrastructure**

      The second category in the literature is computational infrastructure and is the category of Multi-level Mesa. This category has two sub-areas, ABM platforms and coding libraries. ABM platforms are characterized by their own simplified coding language to reduce the barrier of entry for non-programmers. For the ABM platforms there are three which have multi-level models. They are NetLogo[3], SPARK (Simple Platform for Agent-based Representation of Knowledge)[4] and GAMA[5]. NetLogo has an extension dedicated to multi-level models, more accurately meta-models, called LevelSpace. LevelSpace's approach is to link models together (e.g. multi-models) so the dynamics of one can update another. Examples include linking NetLogo's Wolf Sheep Predation model with its Climate Change model where climate impacts grass growth and animal flatulence impacts greenhouse gases and animals whose decision-making function is linked to neural net models (Hjorth A., Head, B., & Wilensky, U., 2015; Hjorth, Weintrop, Brady, & Wilensky, 2016). Based on the taxonomy of this chapter, LevelSpace is infrastructure for the third category, but with OOP this line between connecting models and models with hierarchies and modules is blurry at best. SPARK is a Java-based platform modelled on Netlogo, designed specifically for the use with cell biology. SPARK does not explicitly allow for the formation of hierarchies relying on the implicit nature of object-oriented programming transferred to their coding language to allow the modeler to specify their agents and meta-agents (Solovyev et al., 2010). GAMA, also a Java-based platform, is the only platform which has specific methods for the emergence of new agents formed from lower level agents (Taillandier et al., 2012). Due to applicability of this approach to Multi-level Mesa it is worth looking at GAMA and its methods in more detail.

      GAMA like SPARK, uses the object-oriented nature of Java to embed agents in larger groups. GAMA then proceeds further by providing explicit commands for group formation and algorithms to detect new agents. The commands for group formation include: *capture*, which adds agents to a group, *release*, which removes agents from a group, and *migrate*, which moves agents from one group to another ("Multi-level architecture," n.d.). GAMA also has clustering algorithms embedded within its platform which can be used to specify the use of the capture, release or migrate commands (Taillandier et al., 2012). In addition, GAMA passes properties of Java's objected-oriented language into its GAML language so users can specify different behaviors for groups and their sub-agents and access each agents' respective attributes regardless of the level. GAMA is the only ABM platform which explicitly allows for multi-level architecture within an ABM.

      The second area for computational infrastructure is coding libraries. As each coding library for ABMs (e.g. MASON, Repast, Mesa, FLAME, MaDKit) uses object-oriented programming, each has an implicit ability to have modules and hierarchies. Of

---

[3] https://ccl.northwestern.edu/netlogo/
[4] http://www.pitt.edu/~cirm/spark/
[5] https://gama-platform.github.io/

the existing coding libraries, three identified models consisting of multiple layers Repast[6], FLAME[7], and MaDKit[8] (Haman Tchappi et al., 2018; Morvan, 2013; Smallwood & Holcombe, 2006). Of these three only MaDKit provides explicit infrastructure to support agents operating in multiple levels embedding what it calls the Agent, Group, Role organizational model. Within the MaDKit documentation, this manifests itself in two places, first in the agent who can be assigned to multiple groups and assigned a role in each of these groups. Second, in the network management which maintains the different groups and roles (Michel, Gutknecht, & Ferber, 2017). Although Mason[9] and Repast do not have explicit infrastructure for multi-level models they have built in features which help enable multi-level models. For MASON this includes *Steppable* and anonymous wrappers which allow modelers to group agents together and iterate through them in a schedule and place an agent (or group of agents) in the schedule multiple times (Luke, Cioffi-Revilla, Panait, Sullivan, & Balan, 2005). For Repast it has three features to aid multiple levels, which are also based on scheduling. First, *scheduling annotations* where certain actions are scheduled if a trigger event occurs. Second, *scheduling global behaviors* in which the modeler creates a context which is filled with agents who then are scheduled to behave within that context. Third, *schedule with watcher*, which allows for dynamic scheduling by letting agents know if certain conditions are met so they can execute some action ("Repast Simphony Reference Manual," 2018).

Of the existing platform and coding libraries only two, GAMA and MaDKit, have explicit architecture for developing multiple hierarchies and allowing interaction between them. Although other platforms and coding libraries do not have explicit methods for multiple layers and hierarchies' modelers are able to leverage their object-oriented foundation to develop their own. In addition, MASON and Repast have additional features with their respective scheduler classes which can reduce the cost of integrating modules and hierarchies.

**Category Three: Connecting Models**

The final category is the concept of linking models together to create multi-level ABMs. This category intersects with a much larger field of connecting models and simulations together and are governed by High Level Architecture (HLA) standard of the IEEE (2010). What is significant about coupling models which are part of the same complex system is they will share variables as the various modules in their respective hierarchy update. Unfortunately, this critical dynamic falls outside the IEEE standard (Scerri et al., 2010). Simulating such interdependencies is critical to understanding how these complex interactions may ripple across the entire system. Beyond NetLogo's LevelSpace, this literature review found one effort to deal with this challenge. The paper provides an architecture with two main features to overcome this difficulty, first is a time manager to ensure all the models are synchronized in their sequential management. The second feature is a conflict resolver to determine which model should get access to shared variables first (Scerri et al., 2010). These features go beyond the features Multi-level Mesa will add, but is a dynamic which at some point must be considered. Linking models to simulate the interactions of different modules and hierarchies of a complex system presents new problems not addressed by the common standards of model and simulation coupling.

---

[6] https://repast.github.io/
[7] http://flame.ac.uk/
[8] http://www.madkit.org/
[9] https://cs.gmu.edu/~eclab/projects/mason/#Features

      Multi-level ABMs covers a wide breadth of disciplines and approaches. The natural sciences who are trying to understand the interaction of hierarchies and modules which have evolved over millennia are understandably trying to discover the specifics of those complex interactions. Outside the natural sciences researchers are trying to determine if there are common interaction processes among diverse human societies or develop reliable interaction processes to control fleets of autonomous vehicles. For the majority of ABM libraries and platforms they have relied on the inherent inclusion of modules and hierarchies in object-oriented programming. Only GAMA and MaDKit have explicitly included functionalities for modules and hierarchies, while MASON and Repast have elements in their schedulers which can implicitly aid more complex interactions between modules and hierarchies of agents. The literature review showed that modules and hierarchies are an implicit part of the OOP languages on which ABMs are built. Although modules and hierarchies are an implicit part of OOP languages managing the complex interaction of agents impacting higher level agents and vice versa and the ability for agents to be active in different modules or change from one group to another presents significant management challenges which Multi-level Mesa seeks to mitigate for ABM practitioners.

## **Multi-level Mesa Approach**

      Multi-level Mesa's approach is driven by the concept of complex systems as adaptive networks. The core data structure of Multi-level Mesa is a network graph using Python's NetworkX library (Hagberg, Schult, & Swart, 2008). This approach extends existing multi-level approaches as well as exploiting the OOP nature of Python. The most similar approaches are GAMA and MaDKit, GAMA incorporates clustering algorithms as an additional method of determining if agents are in the same group (Taillandier et al., 2012). Multi-level Mesa, leveraging NetworkX clustering algorithms allows for the same dynamic, while also allowing users to specify when a module may form or activate based on link type or a value associated with a link type. In MaDKit, the user must specify the use of Agent, Group and Role to manage which agent is doing what in which group (Michel et al., 2017). Multi-level Mesa extends this approach by incorporating a dynamic network. Instead of specifying specific groups and roles, as connections between agents change through the dynamics of the ABMs, new modules (groups) can form or dissolve and new behaviors can activate or lay dormant. This approach allows for neutral networks to exist within any model where certain behaviors may only emerge under specific conditions and are not previously seen. Multi-level Mesa goes beyond existing approaches by creating a greater synergy between network science and ABMs, the

interaction of agents produces a dynamic network, which in turn alters the behavior of the agents. The remainder of this section will discuss the specifics of Multi-level Mesa's implementation prior to discussing the results of the Sugarscape ABM used to develop Mult-level Mesa.

**The Multi-Level Mesa Library**

      Multi-level Mesa has three main components. First, a collection of managers which tracks the agents, the modules of agents (groups), the network of agents, agents who belong to an existing group, and the schedule. Second, a series of functions which provides the user different options to form groups or dissolve them. Third, a group class which allows for the inclusion of different group policies, manages the behavior and status of the group, and implicitly produces hierarchies within the complex system. (Figure 3-1)

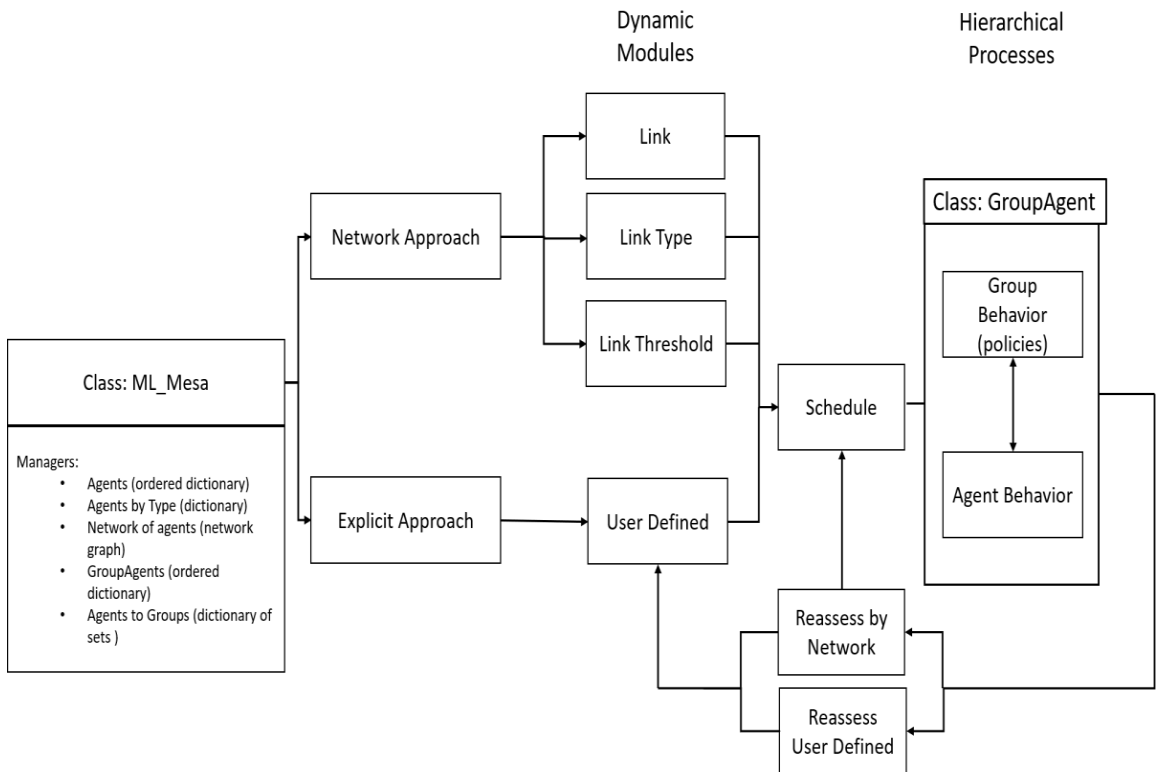

**Figure 3-1: Multi-level Mesa Schematic**

      Multi-level Mesa is available on GitHub at https://github.com/tpike3/multilevel_mesa and is also part of the Python Package Index and can be installed with the *pip install multilevel_mesa.*

***Creating an Multilevel_Mesa Instance and the Multilevel_Mesa Managers***
      Creating an instance of Multi-level Mesa requires no parameters, and initiates one attribute and six managers (Box 3-1). The ML Mesa does have two keyword parameters,

MultiLevel_Mesa.min_for_group and MultiLevel_Mesa.group_net. MultiLevel_Mesa.min_for_group tells the instance the minimum number of agents which must be in a group. The min_for_group parameter has a default setting of two. The second key word parameter is MultiLevel_Mesa.group_to_net, this parameter takes a Boolean and is defaulted to *False.* If a group is added to the network, this tells the Multi-level Mesa instance that the group as an agent can link with other nodes. User specified behavior can then dictate the complexity of these interactions, groups to agents, groups to groups and groups forming groups of groups. The one attribute of Multi-level Mesa is id_counter, which allows for unique_ids to be generated for groups. The six managers are (1) MultiLevel_Mesa._agents which is an ordered dictionary (a hash-table consisting of a key:value pair) that holds every agent added to the instance. This manager is critical to maintain the most granular dictionary possible of all agents and mimics the _agents dictionary found in Mesa. (2) MultiLevel_Mesa.net is an instance of a NetworkX graph. This feature provides the critical structure for tracking and managing agents and groups. (3) MultiLevel_Mesa.agents_by_type uses a dictionary of dictionaries to track agents by type. This feature allows for faster reference of specific types of agents when manipulating groups or schedules. (4) MultiLevel_Mesa.schedule replaces the Mesa schedule and is an ordered dictionary which manages the agents and when they execute a step function. (5) MultiLevel_Mesa.groups is an ordered dictionary and tracks the groups within the model performing the same function of tracking groups as the agents ordered dictionary. (6) MultiLevel_Mesa.reverse_groups is a dictionary of dictionaries of sets. The first dictionary key is the agent id, while the second is group types (link and link values) and the set is the group ids to which the agent belongs in those group types. This structure is necessary to ensure duplicate groups are not created or that an agent is added to an existing group instead of creating a new one. The use of sets also helps expedite computation by using set operations to evaluate if a group should be formed or agents added to an existing group.

> 1. _agents: Ordered Dictionary
> 2. net : NetworkX Undirected Graph
> 3. agents_by_type : Dictionary of Dictionaries
> 4. schedule : Ordered Dictionary
> 5. groups : Ordered Dictionary
> 6. reverse_groups : Dictionary of Dictionaries of Sets

**Box 3-1: Multi-level Mesa Managers and Data Types**

**The Mutli-level Mesa Functions**

As shown in figure 1, Multi-level Mesa has two primary approaches for facilitating a multi-level ABM, an explicit approach and a network approach. Within these two approaches, Multi-level Mesa turns the desired agents into a bilateral link list which form the groups. Each input of agents is transformed into a network edge which forms the groups or adds agents to an existing group. The use of links is also used to disband groups or remove agents from the group. These functions then create a more dynamic schedule with modules of agent within hierarchies.

*Forming and Dissolving Groups*

*User Defined Formation Process: MultiLevel_Mesa.form_group*
  The formation function of the explicit approach is MultiLevel_Mesa.form_group and takes a user defined process which must generate a list of bilaterally connected agents (Box 3-2). This approach can be computationally expensive, but is necessary to allow for the accurate recreation of the network. As dictionaries (e.g. the schedule) cannot be manipulated during iteration users must use a *yield* versus the more common *return* operator to pass the list of agents to the MultiLevel_Mesa.form_group function.

> def form_group(self, process, *args, determine_id = 'default', double = False,\
>    policy = None, group_type = None,  **kwargs):

**Box 3-2: MultiLevel_Mesa.form_meta function**

  The MultiLevel_Mesa.form_group function requires one parameter which is the user specified process which determines whether or not an agent should be in a group with other agents. The *args and **kwargs allows the user to pass in the parameters for this process. The determine_id parameters ensures each group gets a unique id. If *default* it will simply append a number based on the *id_counter* attribute to the string 'group'. For the user to pass in an id he or she must *yield* the id as the first element of a tuple generated from the *yield* operator from the user defined process. Users must choose this id carefully as the id is used in the set operations to merge groups. The double parameter takes a Boolean value and is defaulted to *False*. If *True* the agent will remain in the schedule as an independent entity and be added as part of the group, while if *False* the agent is removed. This feature is to provide users maximum flexibility for agent scheduling and group processes. The policy parameter passes in the step processes for the group, which can consist of only internal processes or can consist of group processes and then execute the individual agent processes.  The group_type parameter takes a string and allows the user to specify different types of groups so an agent can belong to different types of group such as 'family' and 'firm'.

*User Defined Dissolution Process: MultiLevel_Mesa.reassess_group*
  The dissolution function for the explicit approach (although it can be used interchangeably with the network approach) is MultiLevel_Mesa.reassess_group (Box 3-3). This function iterates through each group and then uses the user defined process to assess whether or not an agent should still belong to the group. Similar to the MultiLevel_Mesa.form_group this function requires a *yield* to provide the list of agents which should be removed and then proceeds to remove those agents while updating the appropriate managers. This function also ensures if the group fails to have a certain number of agents within the group that the group will be removed. This minimum number of agents is the min_for_group attribute of the Multi-level Mesa instance and has a default setting of two.

```
def reassess_group(self, process, *args, reintroduce = True,
group_type = None, **kwargs):
```

**Box 3-3: MultiLevel_Mesa.reassess_group function**

The MultiLevel_Mesa.reassess_group function requires one parameter, which is the process defined by the user for assessing whether or not the agent should remain within the group. The function also has a reintroduce parameter which takes a Boolean value and is defaulted to *True*. This parameter tells the function whether or not to reintroduce the removed agents back into the schedule.

*Network Defined Formation: MultiLevel_Mesa.net_group*
The formation function of the network approach is MultiLevel_Mesa.net_group (Box 3-4) and uses an undirected NetworkX graph object to assess what agents should form groups. With an undirected graph and as indicated in figure one, there are three possibilities for assessing whether or not linked agents should be in the same group. First, by whether or not a link exists between the agents. Second, if a specific type of link exists (e.g. friend, enemy). Third, if a link exists which has reached a certain value. For example, in the Sugarscape model discussed in the next section, one version forms a group if an agent and landscape cell are linked, in another version, the agents form a group if they have 10 or more trades between them.

Although, NetworkX also offers the possibility of directed graphs and multi-graphs these options were not used for simplicity sake and because the dynamics of ABMs can account for the main aspects of these features. As NetworkX uses a dictionary structure to capture nodes and links, a multi-graph can be easily simulated by adding more link types along the edge, so a link may have the dictionary keys {family, tribe, job...} allowing for a link with multiple types similar to a multi-graph. The directed graph dynamic can also be achieved through agent interactions as the link attributes can dictate the direction of flow based on agent attributes and behaviors. The one cost is users cannot use the multi-graph and directed graph network evaluation functions in NetworkX. Using an undirected graph provides a leaner, more easily understood approach without loss of network dynamics.

```
def net_group(self, link_type = None, link_value = None,\
              double = False, policy = None):
```

**Box 3-4: MultiLevel_Mesa.net_schedule function**

The MultiLevel_Mesa.net_group function requires no parameters and will default to whether or not a link exists or not between agents. As the group is formed purely based on the links between agents, no *args or **kwargs arguments are required. As the net_group function has no process passed in there is no way to specify a group id, the function uses the default "group" if groups are forming based on the presence of a link, the link_type is not the default *None* or the link_type_link_value, plus a number from the MultiLevel_Mesa.id_counter attribute. If users decided they would like to pass in processes to provide a unique id for groups this could be added in future versions, but was not included in this version as it did not add anything substantive to the Multi-level Mesa dynamics. The link_type function allows the user to pass in what link key value should link agents together. The link_type can then be further specified with the link_value criteria. These values are also used as the dictionary keys in the MultiLevel_Mesa.reverse_groups manager. The link_value can either be a string to further classify the type of link, for example *family: friendly* or *family: angry_teenager* or it can be a value such as will be seen in the Sugarscape model *trades: 10* (number of trades between agents), which in this case tracks a type of interaction between agents. As net_group is an additive process the value is assumed to be a threshold of greater than or equal to a value. The network can then be updated and evaluated through the other processes in the ABM using NetworkX object manipulation functions. For convenience, MultiLevel_Mesa also has MultiLevel_Mesa.add_links and MultiLevel_Mesa.remove_links functions. These functions take a list of agents, combines them in to a list of fully connected tuples and then adds or removes the links.

*Network Defined Dissolution: MultiLevel_Mesa.reassess_net_group*

The MultiLevel_Mesa.reassess_net_group (Box 3-5) uses the same taxonomy of options as MultiLevel_Mesa.net_group. First, an agent can be removed based on the presence of a link, the presence of a specific link type and finally the presence of a specific link value. The function will also check to ensure the meta-agent still has the minimum number of agents to remain a group which is defaulted to two with the MultiLevel_Mesa.min_for_group attribute.

```
def reassess_net_group(self, link_type = None, link_value = None)
```

**Box 3-5: MultiLevel_Mesa.reassess_net_group function**

The dissolution function similar to the formation function requires no parameters and will default to determining if there is a link or not. The user can also specify link types which cause agents to be removed or link values, which can again be either strings or numbers. However, as this function is not additive, the agent will be removed if the value is less than or equal to the user specified value.

*A Note on Formation Precedence*

A critical point for users to understand is agents belong to the first group with which they form. If an agent is not part of a group and meets the user given criteria it will then be added to the first group evaluated by function based on the specified user dynamics or randomly ordered dictionary of the NetworkX link dictionary. If both agents belong to a group the link between them at the agent level will remain in place. This approach was adopted because the dynamics of how agents should be integrated into groups is specific to the user's model. This approach, therefore, defaults to the first group joined which is consistent with human biases (Heuer, 1999; Pratkanis & Aronson, 2002). Appreciating how this dynamic works will allow users to leverage the other functions to specify group precedence.

*Schedule Functions*

As MultiLevel_Mesa replaces the normal schedule function of Mesa, it must also have the basic scheduling functions (Box 3-6). These are the add and remove functions, which remain at the individual agent level but have a higher degree of complexity as agents must be kept in multiple managers to ensure agents are being properly 'stepped' in the schedule or removed if the agent 'dies'. Multi-level Mesa also replaces Mesa's step function. Its primary schedule is random activation, but this can be turned off for an ordered activation and a staged activation can be executed through the agent_type manager. A future extension of MultiLevel_Mesa would be to store different schedules based on different network configurations. This would save computation time so specific agent schedules would be created less often. For example, if one was recreating daily life of a population and the night and morning hours used one configuration, while the daytime hours would use a different configuration, each calling different behavior routines for the agents.

```
def add(self, agent, schedule = True, net = True)
def remove(self, agent):
def step(self, shuffled = True, by_type = False, const_update = False)
```

**Box 3-5: MultiLevel_Mesa. Schedule functions**

Similar to Mesa, the MultiLevel_Mesa.add function requires an agent object. It also has two keyword parameters which take Boolean parameters each with a default value of *True*. Keyword parameter *schedule* adds the agent to the schedule. This is an option in case the user begins with a complex network and the agent is already part of a group. The *net* parameter similarly adds the agent to the NetworkX object. This is done in case the user has an agent he or she does not want to be part of the network. For instance, in a Sugarscape model, the grid cells may not need to be a part of the network as what is of concern is the agent's network. The MultiLevel.Mesa.remove function requires an agent object. If invoked this will remove the agent from all managers as applicable. The MultiLevel_Mesa.step function works in a similar way to the Mesa step function, where it iterates through each agent in schedule and executes their step function. Random activation is the default as identified by the keyword parameter *shuffled*. If shuffled is *False* it will follow the order in the ordered dictionary (the order the agents were added). The keyword parameter *by_type* is set to *False* but can take a list of agent types to simulate staged activation. Constant update provides the ability to have specific agent types activated after the more dynamic schedule. For example, an environmental variable which changes at a steady rate for each time step, such as sugar or spice growth in the Sugarscape model.

**The Group Class**

The Group class introduces hierarchy into the ABM. The Group class performs similar functions to Multi-level Mesa or Mesa's time module. The Group class has three managers, which includes a dictionary of the agents which belong to the Group, a dictionary of dictionaries with the agents in the Group by type and a NetworkX graph object of the sub_agents. The Group then has three attributes to make it easier for users to employ the Group. The first attribute is Group.active which is a Boolean value to help users activate and deactivate Groups as necessary. The next two attributes are Group.type and Group.__str__ which both equal "group" and allow the user greater ease in identifying and performing functions on the groups. The final attribute of the Group is its policy object, this object is passed in by the user and provides the Group behavior. The behavior of the Groups and its internal agents is done with two step functions the Group.group_step which calls the policy function and the individual agent step functions, again using a random order, but with the same options of the MultiLevel_Mesa.step function to dictate schedule ordering processes.

> **Attributes:**
>   Group.sub_agents = dictionary
>   Group.agents_by_type = dictionary
>   Group.net = NetworkX graph
>   Group.policy = object of group policies
>   Group.active = status of Group
>
> **Main Functions**:
>   Group.meta_step() = policies to dictate sub_agent behavior
>   Group.step() = sub_agent behaviors

**Box 3-6: Group attributes and functions**

The interaction of the schedule, formation and dissolution of modules of agents, and the ability for hierarchies to exist allows for the easier introduction of these key features of complex systems. The functions can be employed as part of the normal step function, at specific events or at specific intervals. By using a network data structure as the main management structure, Multi-level Mesa is able to integrate the interdependencies and changing dynamics of those interdependencies into ABM management structure providing a new dynamic which goes beyond the current multi-level approaches. With an understanding how the main functions and dynamics of Multi-level Mesa, it is now time to verify and validate the Multi-level Mesa library.

## **Multi-level Mesa and Sugarscape**

Sugarscape was used to verify and validate the functioning of the Multi-level Mesa library.[10] Sugarscape was one of the first ABMs to demonstrate bottom up emergence of system behavior based on the decentralized action of many agents. The specific variation used for Multi-level Mesa is the trade variation in which the landscape has two commodities sugar and spice and the agents must acquire and consume both based on their unique sugar and spice metabolism in order to survive. The agents trade their sugar and spice accumulations based on the amount of sugar and spice they have acquired and their marginal rate of substitution due to their metabolisms (Axtell & Epstein, 1996). This variation offers a great test case for Multi-level Mesa because the results are well founded on economic theory providing clear verification and validation for Multi-level Mesa, as well as providing enough complexity that groups can be introduced in different configurations and with different policies to show the impact of this additional dynamic.

The Multi-level Mesa model uses the base case of a trading environment outlined in the beginning of chapter four of *Growing Artificial Societies*. The landscape is a 50 by 50 torus with each cell given a quantity of sugar and or spice from zero to six (Figure 3-2). There are four mounds, each with a gradient that increases in sugar or spice as one gets closer to a peak. Each grid will regrow one sugar and one spice unit per step until its maximum sugar and spice allotment is reached. There are 200 hundred agents, each instantiated with a vision attribute between one and six which determines how many cells they can see using a Von Neumann neighborhood (four cardinal directions). Each agent is instantiated with a sugar and spice metabolism between one and six, which indicates how much sugar or spice each agent consumes with each step. Each agent is also given an initial endowment of sugar and spice from 25 to 50. On each time step, the schedule iterates through a randomly ordered list of agents and each agent moves to collect more sugar and spice, consume sugar and spice, and trade with agents within their vision. The agents move and trade based on their marginal rate of substitution and in accordance with what their vision allows according to a Von Neumann neighborhood and as calculated in *Growing Artificial Societies* (Axtell & Epstein, 1996). With this model, the different configurations of Multi-level Mesa are tested to both verify and validate its use as a library.

---

[10] All code for this instantiation of Sugarscape can be found at https://github.com/tpike3/SugarScape. Due to the size the results were not included, but the code used to analyze the results was included. This allows any interested parties to run and analyze the code. The results can also be provided upon request.

Testing Multi-level Mesa occurred in three phases, the first phase is showing equivalency between Multi-level Mesa's explicit and network approach and a standard Sugarscape configuration. The second phase is showing equivalency with the formation of groups and the third phase is showing the impact of different group policies on agent behavior. This provides both verification and validation of the Multi-level Mesa library as well as justifying its existence based on the impact of even simple group policies on emergent behavior.

**Phase I: Equivalency Between Multi-level Mesa Approaches and a Standard Approach**

The first phase recreates Sugarscape, specifically the sugar and spice variation described in chapter four of *Growing Artificial Societies* (Axtell & Epstein, 1996), and replicates the output of this standard approach using Multi-level Mesa's explicit and network approach. In the standard approach the schedule randomly orders each agent as they iterate through the movement, consumption and trade functions. This model replicates the key result of the sugar and spice landscape as the price of both sugar and spice moves toward one and the standard deviation of the logarithmic mean of the price moves toward zero, as predicted by economic theory (Axtell & Epstein, 1996). This instantiation of the sugar and spice landscape does not match the trade volume in *Growing Artificial Societies* as the volume total is much less and follows a heavily skewed distribution (Figure 3). This difference is acceptable as the metric for validation is not the amount of trade but rather the trade price (Axtell and Epstein 1996). To ensure the proper functioning of the Multi-level Mesa library these results then needed to be replicated using the network and explicit approaches.

To replicate these results using the explicit approach, the model forms a group with each agent and the landscape cell on which the agent is located. The model then steps forward each group, producing the same set of dynamics as the standard approach. The model then reassesses each group and if the agent has moved disbands the group. Similarly, for the network approach, the model forms a link between the agent and the cell it is on, the model steps through the agent functions and then the link between the agent and cell is reevaluated and removed if the agent is no longer on the cell. These two approaches then replicate the results of price, standard deviation of the logarithmic mean, and trade volume (Figure 3-3). In addition, these three variations were run for 100 runs over 1000 steps mimicking *Growing Artificial Societies* (Axtell & Epstein, 1996). The output of their respective price distributions was not qualitatively different and their surviving number of agents was not statistically different demonstrating the approaches are equivalent versions of the same dynamic. The additional process of creating and destroying the groups added a time cost from the standard variation with a two second addition to the mean for the network approach and a 14 second addition to the mean for the explicit approach (Figure 3-4). These results provide the simplest possible comparison to ensure the Multi-level Mesa approach does not fundamentally alter ABM dynamics. These results then allowed for advancement to phase two introducing groups of trading agents.

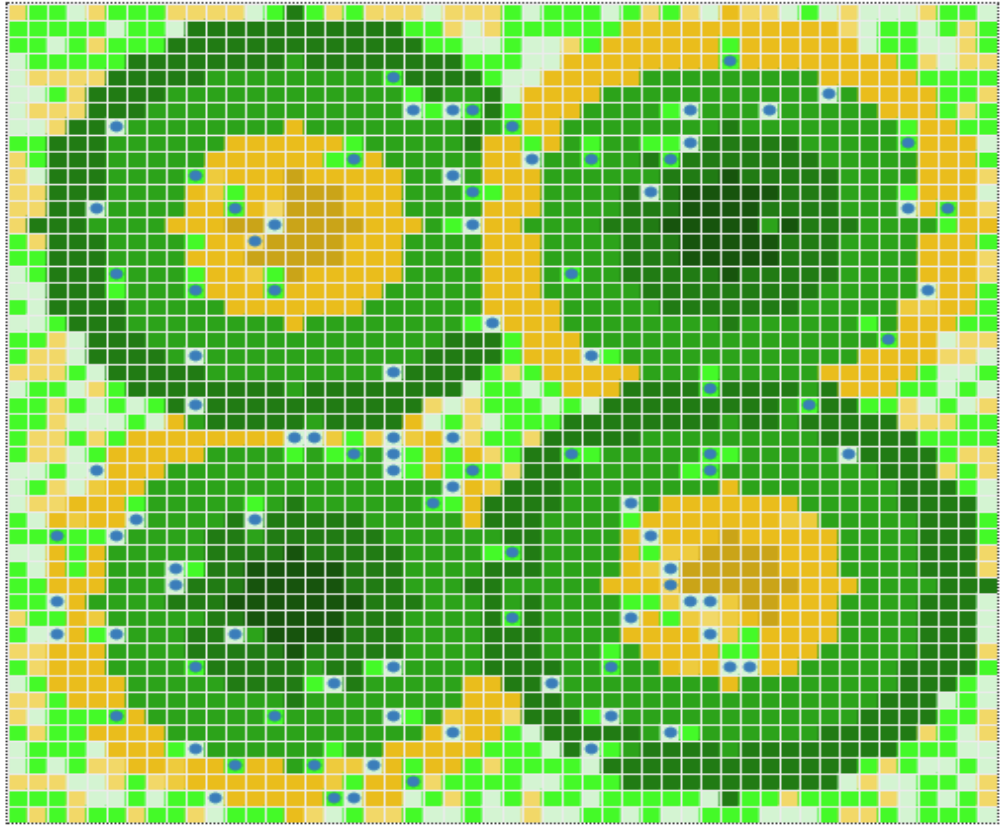

Figure 3-2: Sugar and Spice Landscape, the tan peaks represent more spice while the green peaks represent more sugar, the dots are the agents.

**Phase II: Introducing Groups**

In phase two the use of the Multi-level Mesa library was varied so groups formed if two or more agents reached a threshold of trades. The primary question for this phase was whether or not grouping agents together in the schedule would have any impact on the results. As demonstrated in *Who Goes First? An Examination of the Impact of Activation on Outcome Behavior in Agent-based Models*, activation schemes in ABMs do matter (Comer, 2014). In this case however, agents grouping together and being randomly activated as a group had no significant outcome on the results. To test this, groups were formed at one, five and 10 trades over 1000 steps and 100 runs. The overall results were compared, as well as specific results of the price and trade volume. For each parameter the grouping of agents had no observable impact on the results. Figure 3-5 shows agent configuration, the price, and standard deviation of the logarithmic mean of the price. Figure 3-6 shows the overall price distribution of 100 runs for the explicit and network approach as compared to the standard approach. Although shown for 10 trades, groups which formed at one and five trades showed similar results. The network approach time per run was comparable with the standard approach and actually had a mean of three second less. While the explicit approach incurred a time cost of 15 seconds. The reason for the network approach is comparable is it is able to use NetworkX's dictionary structure to reference specific agent groups, while the explicit approach must iterate through the model's trade dictionary and assess the trade status. Although, grouping agents in the schedule did not have an impact on this particular model this does not mean these results are generalizable. For this study, however, these results provide further verification of the functioning Multi-level Mesa.

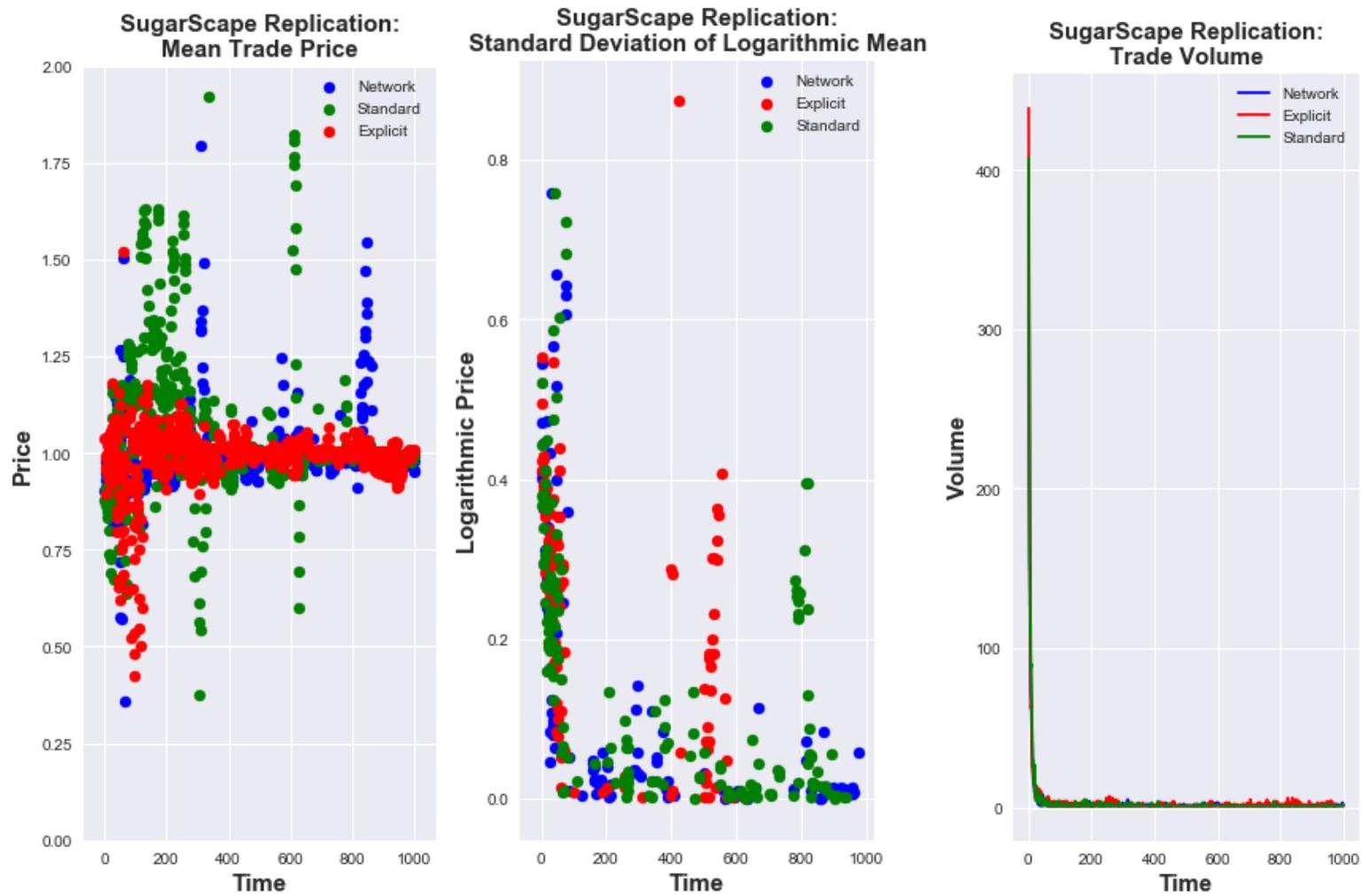

**Figure 3-3: Single Run Results of Price, Standard Deviation of Logarithmic Mean and Trade Volume**

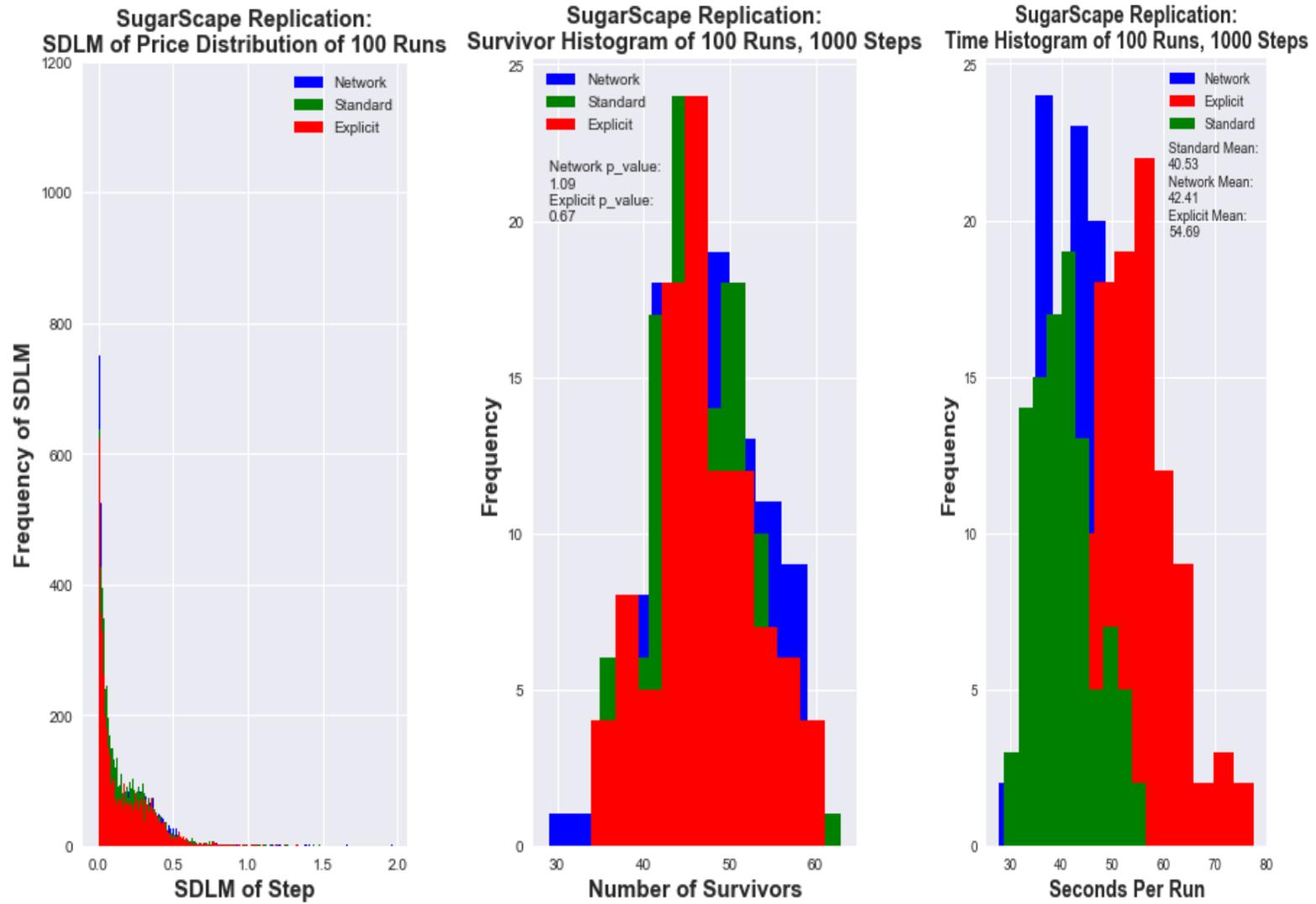

Figure 3-4: Standard Deviation of Logarithmic Mean Price Distribution, Survivor and Time Histograms for 100 runs of 1000 step

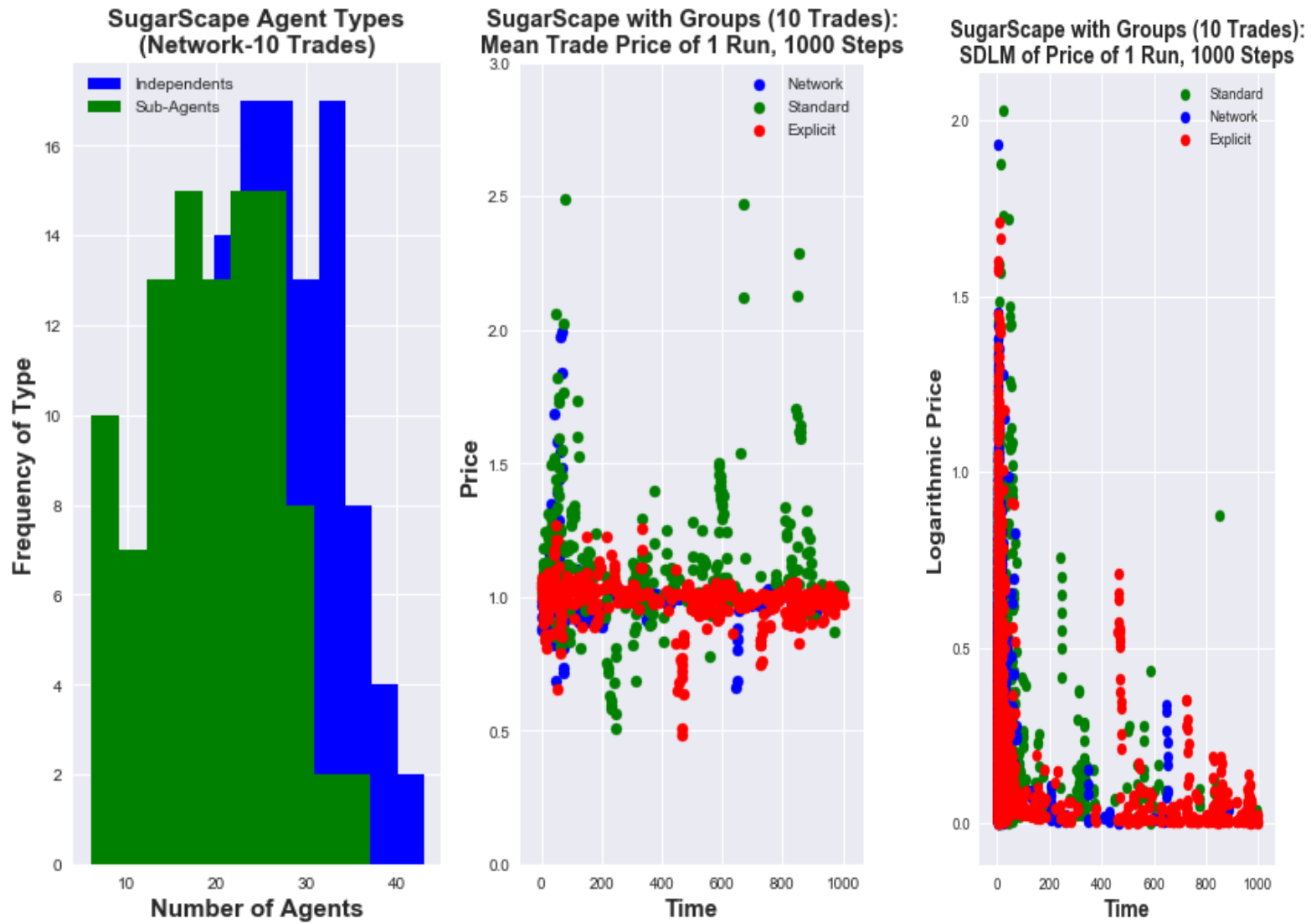

**Figure 3-5: Agent Types for 100 runs, Mean Trade Price for One Run, Standard Deviation of Logarithmic Mean for One Run**

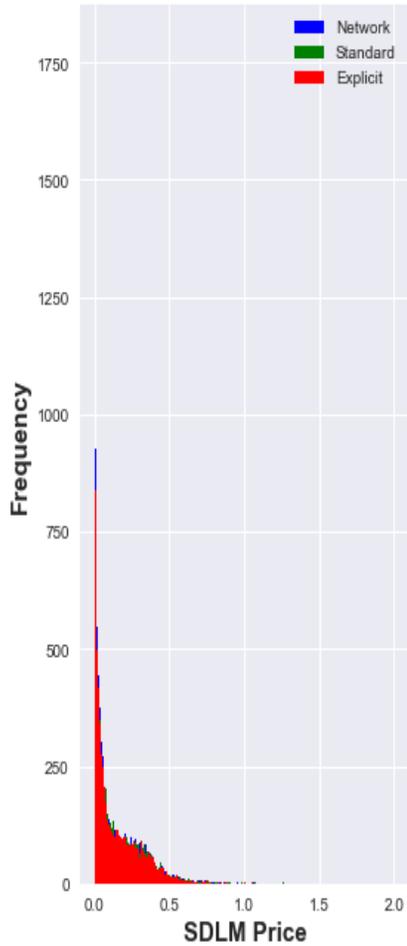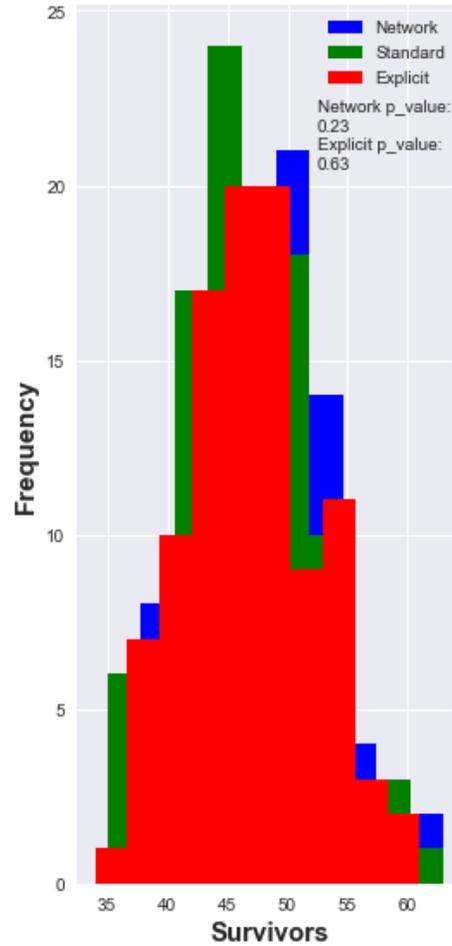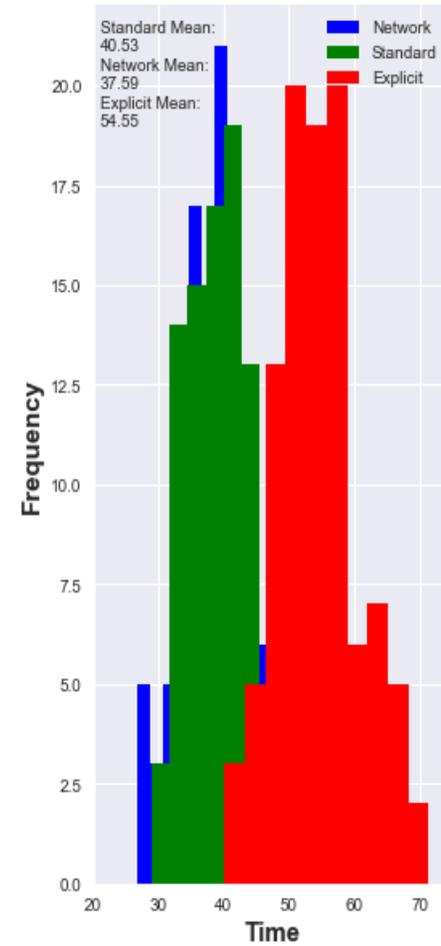

**Figure 3-6: Standard Deviation of the Logarithmic Mean, Survivors and Time Histogram Results of 100 Runs with 10 Trades Forming a Group**

**Phase III: Introducing Group Policy**

The final phase introduced policy into groups. Policy for Multi-level Mesa is understood to be group behavior which alters behavior of the agents in their group. If the agent is not part of the group it will behave differently. To assess the impact of policy three variations were implemented. First, the policy of the group changed the individual agent's behavior. Second, the group resources were available to all, but the agents explored and traded based on their own situation. Third, the groups shared their resources and explored as in the second version, but the groups could form groups with other groups. This tested the ability of Multi-level Mesa to allow multiple levels to emerge endogenously as the group could trade with other groups and form a group consisting of sub-groups and individual agents. Phase three tested the impact of group policy of the emergent behavior of the system.

For the first variation, agents reaching a specified number of trades (i.e. one, five and 10) formed a group and the group agent applied a policy to their behavior which changed how the group's agents moved through the landscape. Once a group agent is formed it is randomly assigned one of three policies. (1) Each agent within the group moves to a new cell as though it has the lowest sugar or spice accumulation in the group. (2) Each agent within the group moves to a new cell as though it has the highest sugar or spice accumulation in the group. (3) Each agent moves to a new cell as though it has the geometric mean of sugar and spice accumulations of the group. It is important to note, the agent's perception was only changed based on accumulation, their metabolism was not changed so the agent explored the environment with their view of which cell provided the best resources based on their metabolism and their respective groups accumulation policy. The agents then continued to consume and trade based on their actual accumulations. These policies had a substantial impact on the outcomes of model (Figure 3-7 and 3-8).

As shown in figure 3-7, the policies prevented the price from moving towards one and the standard deviation of the logarithmic mean from moving towards zero. In addition, and somewhat surprisingly, the policies reduced the number of independent agents as compared with no policy (Figure 3-5). Reducing the number of trades required to form a group further reduced the number of independent agents. This occurred because the agents when reaching a price equilibrium were also reaching a movement equilibrium. The policies of the groups then caused the agents to explore more of the environment increasing the number of agents in contact and trading with each other, and reducing the number of independent agents. The policies however, reduced the survival rate of the agents and so if they had a choice, it would not be in the agent's interest to be part of a group (Figure 3-8). Other variations in which the agents searched based on the total group accumulations and could trade with everyone in their group regardless of distance and vision, also resulted in lower survival rates. From these results, one can conclude in Sugarscape searching the landscape based on someone else's situation is sub-optimal. As will be seen in the next variation however, sharing one's resources, and searching and trading based one's own metabolism can lead to much greater survival rates.

Comparing the times between model approaches the network approach was slightly faster, while the explicit approach incurring approximately 47 seconds per run (Figure 3-8). The reason for the explicit is the additional iterations the explicit approach must do in order to assess the groups agents each step, while the network approach is able to reference the dictionary structure of the NetworkX object and its links. Comparing the mean for Network and Standard variations with and without policy shows the these two versions are effectively comparable with time, while other processes in the computer are impacting the exact results.

The next variation for the group agents consolidated the accumulations of each agents to create a common resource available to all group agents. Agents, however, would interpret this accumulation through their own metabolism for trade and movement. This variation was then further explored by the Multi-level_Mesa instance parameter for adding the group to the network to allow group agents to form links with other group agents, creating multiple levels. Agents trade and form a group, then these groups trade and form a group and so on. These variations were only instantiated using the network approach but can be done in the explicit approach as well.

The results continued to demonstrate that group policy changes the emergent behavior of the system. For both one level and multiple levels of agents accessing and consuming groups resources, while exploring their environment based on their metabolism, there was a change in the qualitative shape of the price distribution curve and increases to the survival of the population (Figure 3-9). The inclusion of multiple levels had no impact on the results, which makes sense as the behavior is moving and trading is at the agent level and the groups only provide a common accumulation. Reducing the number of trades (e.g. from 10 to five) did not change the shape of price distribution but it did change the height, increasing the center peak by approximately 400. This increase in trade frequency can be attributed to the dynamic that the sooner the agents are able to join a group the higher their chance of survival. Faster group formations had higher survival rates.  These survival rates then changed the time dynamic of each run, more surviving agents resulted in longer run times. The majority of computation cost being in the agent also produced the counter intuitive results that the multi-level time mean was less than the mean time for one level runs. As the multi-level survival mean was slightly less than that of the one level survival mean (92.83 vs. 94.35) it was therefore slightly faster but not statistically different (p-value of .115) (Figure 3-9).

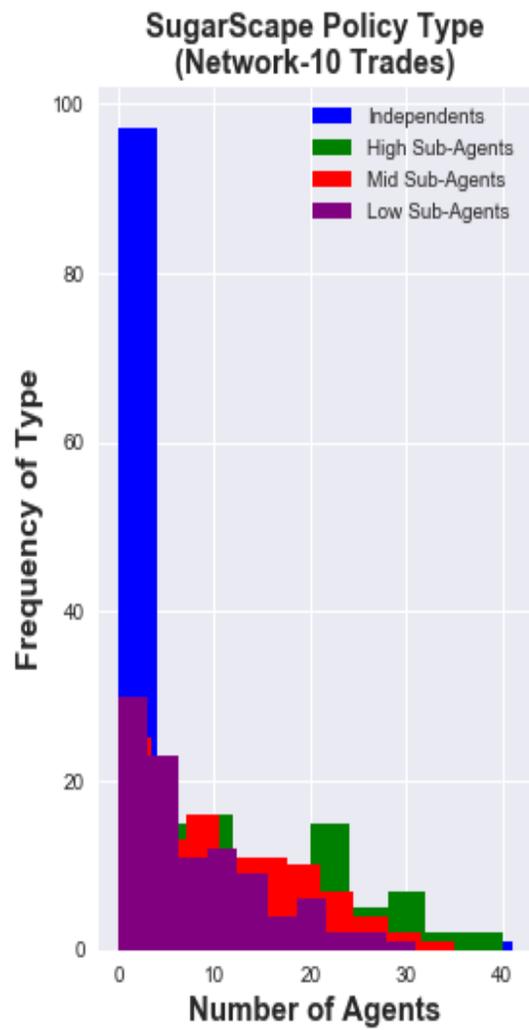 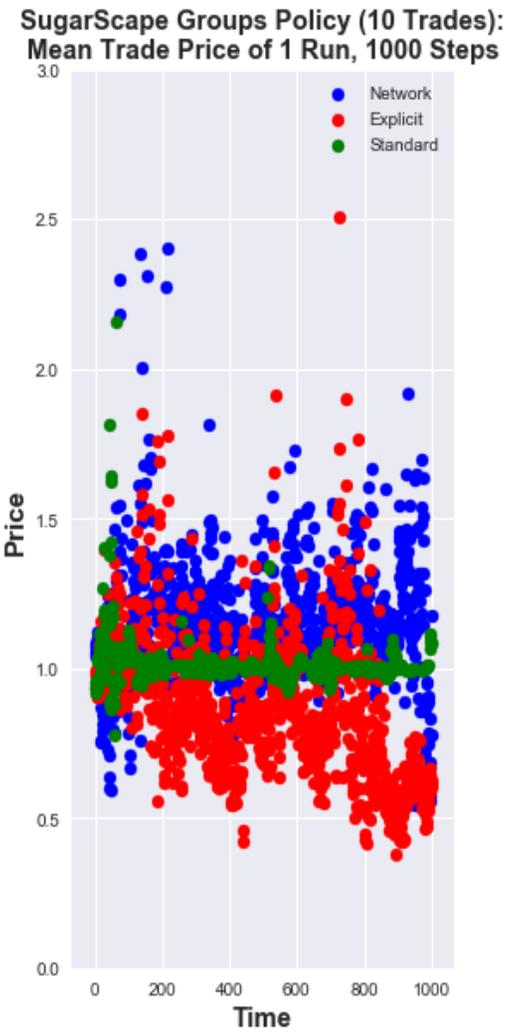 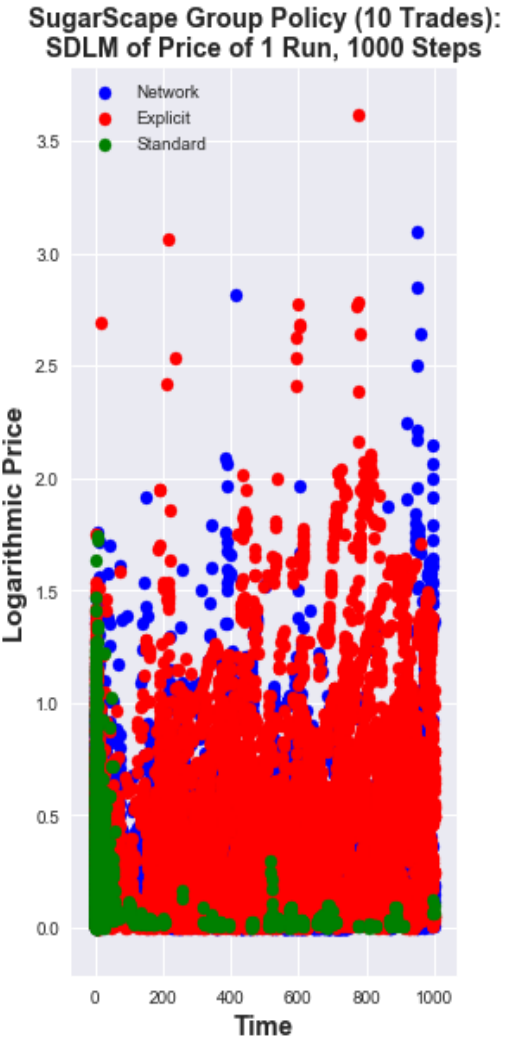

**Figure 3-7: Impact of Policies on Groups, Mean Price and Standard Deviation of the Logarithmic Mean**

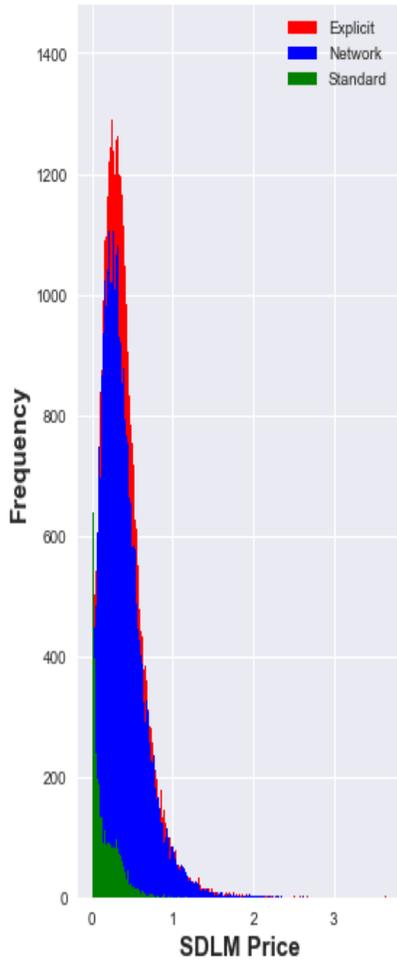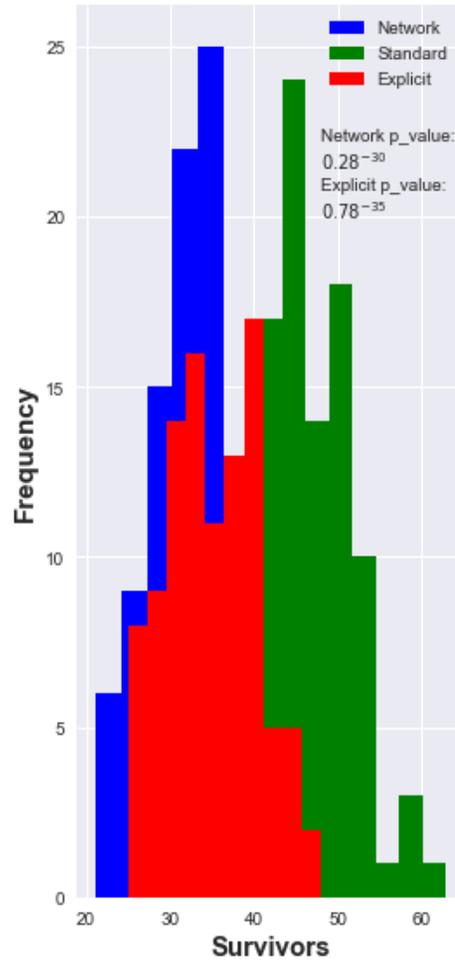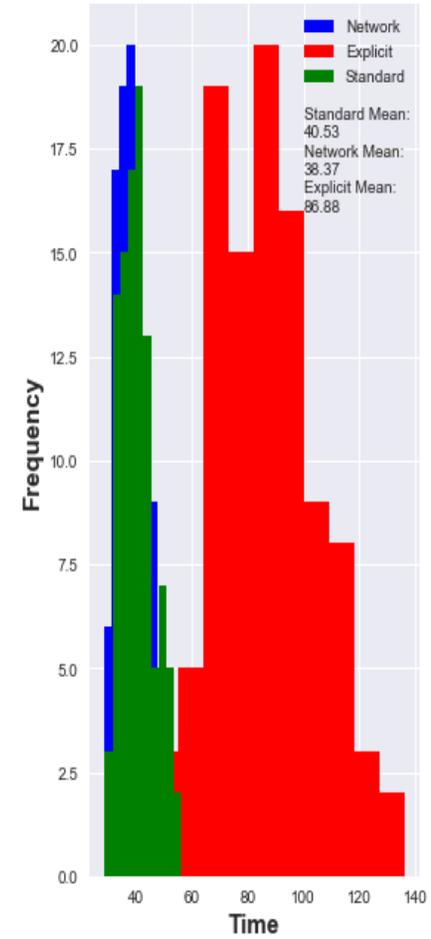

**Figure 3-8: Standard Deviation of Logarithmic Mean, Survivors and Time Histograms for Groups with Policies for 100 Runs, 1000 Steps**

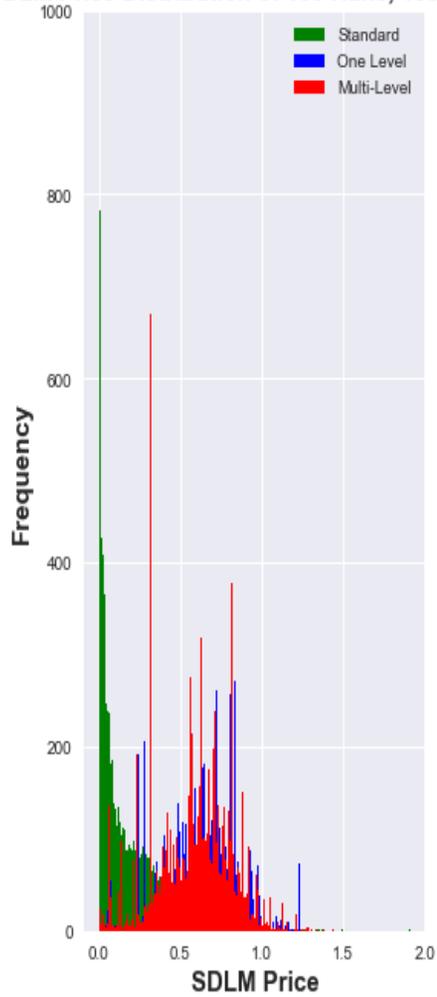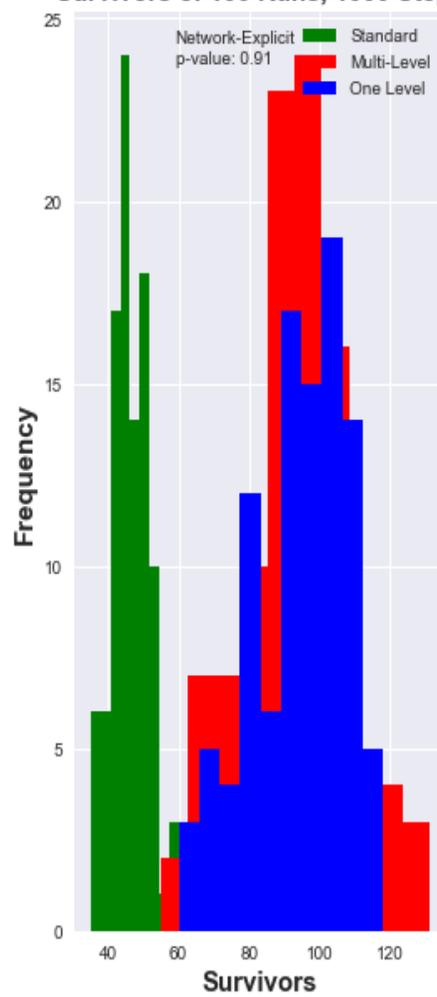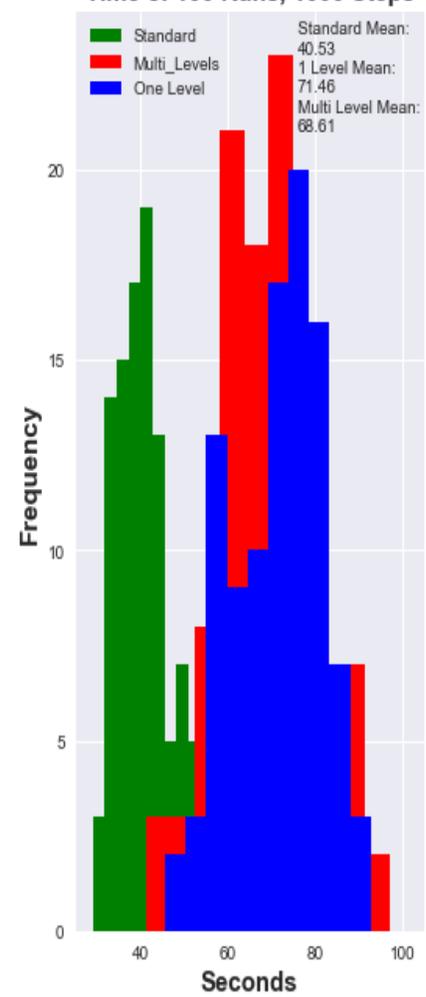

**Figure 3-9: Price Distribution, Survivor and Time Histograms for 100 Runs, with 1000 Steps of Agent Groups whose Accumulations are a Common Resource**

These results continued to demonstrate group policy matters and demonstrated Multi-level Mesa's ability to recursively form multiple levels of groups.

Additional variations were attempted to further verify the group formation of Multi-level Mesa. When vision or agent density was increased, as expected, the agents tend to coalesce towards one super group with levels of groups within them, for this model up to four (one super group, two levels of intermediate groups, and individual agents). In addition, the order of action in the group was varied. In one version, the agents within the
group all moved and collected, before eating and trading. In another variation each agent moved, collected, ate and traded in a random order. This had no noticeable impact on the results.

## **Summary**

The sugar and spice variation of Sugarscape served as an excellent dynamic to both verify and validate Multi-level Mesa. Through Sugarscape, Multi-level Mesa successfully showed that first it 'does no harm' and can successfully replicate the key dynamics of the sugar and spice trading model. Second, it showed that agents forming groups with a policy to direct their agent's behavior, does impact the emergent behavior of the system. Both the shape of the price distribution and survival rates of the agents were impacted based on the policy. Finally, Multi-level Mesa demonstrated that it can create multiple levels of groups through recursion. These results were shown with a new way to conceptualize agent-based scheduling, networks. Dynamically changing agent activity based on complex networks offers new modelling possibilities which are consistent with other complex adaptive systems such a brain activity and genetic networks. Multi-level Mesa will help reduce the barrier of entry to analysts, planners and decision makers, while increasing their ability to develop models of the complex societies they are trying to influence. These models will help them conduct virtual experiments with complex population networks in pursuit of more effective policy with less resources.

*Imaging: Macro to Nano, 2006.* (pp. 816–819). Arlington, Virginia, USA: IEEE. https://doi.org/10.1109/ISBI.2006.1625043

Solovyev, A., Mikheev, M., Zhou, L., Dutta-Moscato, J., Ziraldo, C., An, G., … Mi, Q. (2010). SPARK: A Framework for Multi-Scale Agent-Based Biomedical Modeling. *International Journal of Agent Technologies and Systems*, *2*(3), 18–30. https://doi.org/10.4018/jats.2010070102

Soyez, J.-B., Morvan, G., Dupont, D., & Merzouki, R. (2013). A Methodology to Engineer and Validate Dynamic Multi-level Multi-agent Based Simulations. In F. Giardini & F. Amblard (Eds.), *Multi-Agent-Based Simulation XIII* (Vol. 7838, pp. 130–142). Berlin, Heidelberg: Springer Berlin Heidelberg. https://doi.org/10.1007/978-3-642-38859-0_10

Taillandier, P., Vo, D.-A., Amouroux, E., & Drogoul, A. (2012). GAMA: A Simulation Platform That Integrates Geographical Information Data, Agent-Based Modeling and Multi-scale Control. In N. Desai, A. Liu, & M. Winikoff (Eds.), *Principles and Practice of Multi-Agent Systems* (Vol. 7057, pp. 242–258). Berlin, Heidelberg: Springer Berlin Heidelberg. https://doi.org/10.1007/978-3-642-25920-3_17